\documentclass[manuscript,screen]{acmart}

\AtBeginDocument{%
  }

\begin{document}

\title{Improved Focus on Hard Samples for Lung Nodule Detection}


\author{Yujiang Chen}
\email{1254617738@qq.com}
\orcid{0009-0008-3087-0564}
\affiliation{%
  \institution{University of Electronic Science and Technology of China}
  \streetaddress{No.2006, Xiyuan Ave, West Hi-Tech Zone}
  \city{Chengdu}
  \state{Sichuan}
  \country{China}
  \postcode{611731}
}

\author{Mei Xie*}
\orcid{0000-0001-5605-8867}
\affiliation{%
  \institution{University of Electronic Science and Technology of China}
  \streetaddress{No.2006, Xiyuan Ave, West Hi-Tech Zone}
  \city{Chengdu}
  \state{Sichuan}
  \country{China}
  \postcode{611731}
}


\begin{abstract}
Recently, lung nodule detection methods based on deep learning have shown excellent performance in the medical image processing field. Considering that only a few public lung datasets are available and lung nodules are more difficult to detect in CT images than in natural images, the existing methods face many bottlenecks when detecting lung nodules, especially hard ones in CT images. In order to solve these problems, we plan to enhance the focus of our network. In this work, we present an improved detection network that pays more attention to hard samples and datasets to deal with lung nodules by introducing deformable convolution and self-paced learning. Experiments on the LUNA16 dataset demonstrate the effectiveness of our proposed components and show that our method has reached competitive performance.
\end{abstract}



\keywords{Lung nodule detection, Focus, Deformable convolution, Self-paced Learning}


\maketitle

\section{Introduction}
Lung cancer is one of the most common diseases, which is a serious threat to human health \cite{b1}. According to the Global Cancer Statistics 2022 \cite{b2}, lung cancer accounted for 18$\%$ among cancer deaths (ranking first), where the data is 21.5$\%$ in males and 13.7$\%$ in females. Medical experts believe that early diagnosis of lung cancer is an effective way to reduce mortality \cite{b3}. In recent years, diagnostic methods based on CT images, which contain detailed information about lung nodules, have been widely used clinically \cite{b4}. Therefore, radiologists can make an accurate diagnosis by reading CT images. However, examining 3D lung voxels on 2D CT images is challenging for the naked eye. Due to the lack of professional and experienced radiologists, manual diagnostic methods cannot meet the needs of clinical applications. In order to detect lung nodules in complex lung environments and reduce misjudgment, Computer-Aided Detection (CAD) systems were proposed as the radiologists' assistant \cite{b5}.

Traditional CAD systems use handcrafted features for detection tasks \cite{b6,b7,b8}. However, with the exponentially increasing number of CT images, selecting suitable features from the feature space becomes a rigorous task. Due to the continuous development of deep learning, especially that of CNN \cite{b9,b10,b11}, CNN-based CAD systems have shown remarkably outstanding performance in the field of medicine to diagnose diseases, and they are proven to be practical and valuable \cite{b12,b15,b16,b18,b20}. CNN-based CAD systems can extract features automatically by learning, but the training of the model mainly depends on datasets. For example, Setio et al. \cite{b15} fused six diagonal views processed by ConvNets stream as inputs to enhance feature extraction of small lung nodules. Hamidian et al. \cite{b18} trained a 3D CNN on difficult negative samples generated by an FCN to resolve the false detection. Cao et al. \cite{b20} proposed a TSCNN contained residual dense mechanism and a sampling strategy to further improve the feature extraction. However, there are still two problems for lung CT image processing. The first one is that lung nodules only occupy small areas in CT images and are different in shape. Thus, we need to focus on these areas and strengthen the feature extraction. The second one is that hard samples could cause the reduction of the accuracy of CNN models, such as lung nodules adhering to other lung tissues. We need to pay more attention to the learning of hard samples in the training process. In addition, many researchers have introduced various tricks to improve the performance of CNN-based CAD systems for lung nodule detection. However, these methods did not concentrate on the focus of CNN.

\begin{figure}
\centering
\includegraphics[width=0.55\linewidth]{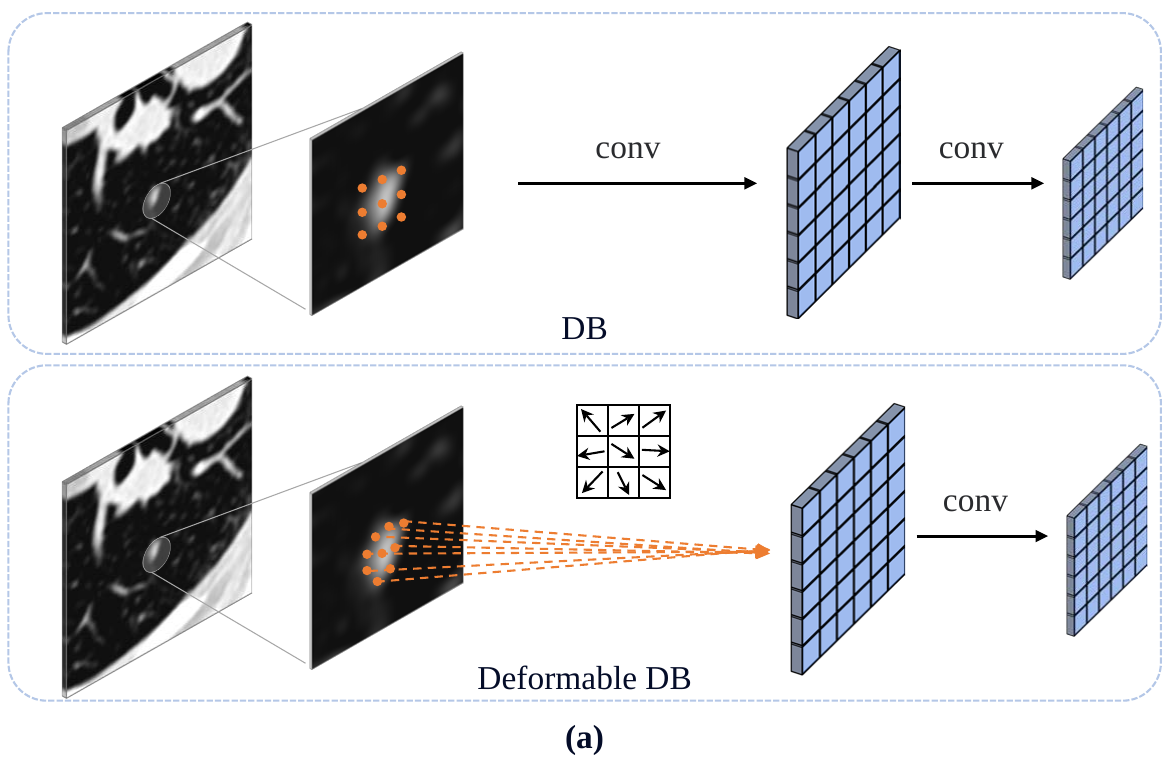}
\includegraphics[width=0.55\linewidth]{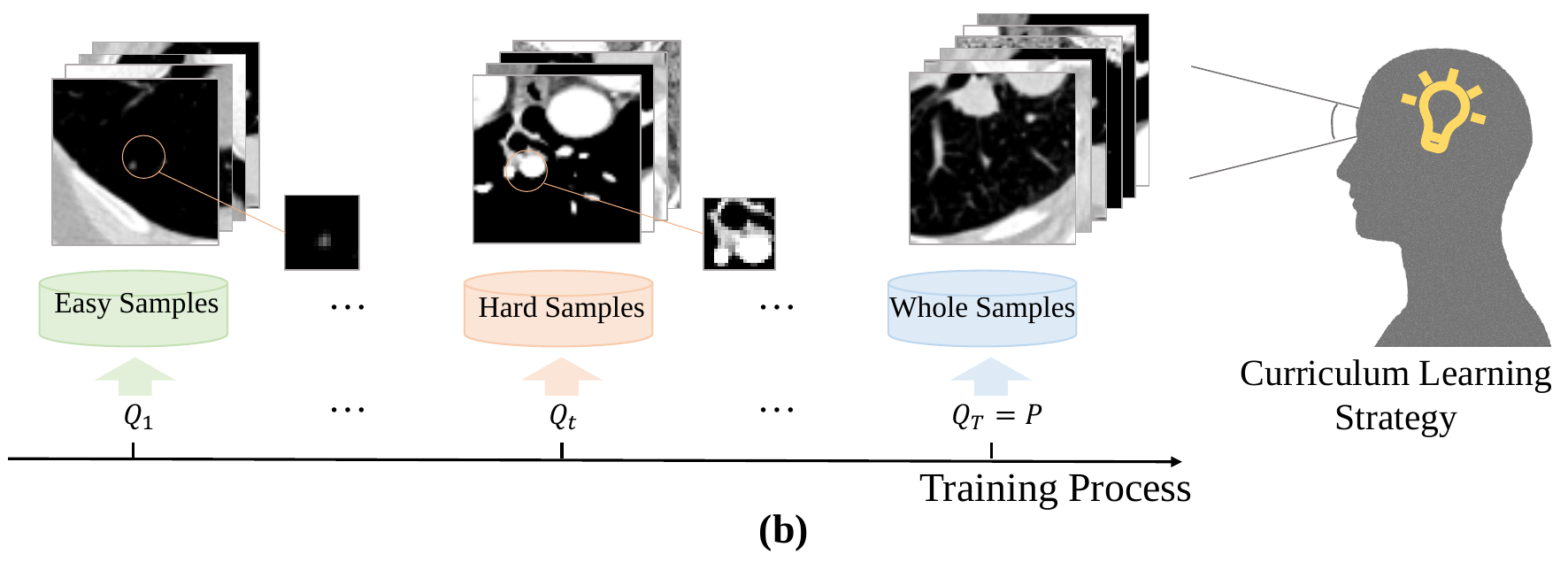}
\caption{(a) shows an indication of fixed receptive fields in standard convolution (up) and adaptive receptive fields in deformable convolution (down). In (b), the left of each image triplet shows the sampling location of a $3\times 3$ filter on the preceding feature map of lung nodules. Curriculum learning strategy simulates the learning process of human beings for new knowledge, dividing the training process into several stages from the simple to the difficult,  and eventually mastering it.}
\Description{The deformable convolution and the curriculum learning strategy}
\label{fig1}
\end{figure}

In this work, we further improve the feature extraction and training strategy to focus on the hard samples and obtain better performance. Firstly, we improve the dense block (DB) based on deformable convolution \cite{b22}, as shown in Fig. \ref{fig1} (a), which can adjust the receptive field to better cover the region of interest. We use the designed deformable DB to enhance the focus of lung nodule detection networks, and verify that the new block can improve the model performance with a bit of time consumption. Secondly, we propose to apply a learning strategy inspired by the curriculum learning \cite{b23} to simulate the learning process of a beginner radiologist, dividing the training process into two stages, as shown in Fig. \ref{fig1} (b). We let our model learn simple samples,  and then we gradually increase the difficulty so that the model will focus on the hard samples \cite{b24,b25}. At last, our model becomes more experienced in detecting hard lung nodules. In this method, we can train an outstanding model that has better performance in lung nodule detection tasks.

\section{Methodology}
\subsection{Baseline}

The detection of lung nodules usually involves two stages where one stage is for the detection of lung nodule candidates and the other is for the reduction of false positive lung nodules \cite{b26}. In this section, we review the architecture of TSCNN (Two-Stage Convolutional Neural Networks) \cite{b20}. An improved U-Net architecture with the residual dense mechanism was presented as the first stage of TSCNN to better detect lung nodule candidates. Meanwhile, an offline hard mining strategy was used to improve the recall rate. The second stage was based on the multi-branch ensemble learning (MBEL) architecture that contains three 3D CNN network models to reduce the false positive lung nodules. At the same time, the proposed dual pooling approach that combines central pooling and central cropping can extract multi-scale features. In addition, two-phase prediction and random mask data augmentation methods are used to ensure a high recall rate.

\subsection{Deformable Dense Block}

Although the residual dense block (RDB) has reached great performances on small-size lung nodules, the models still cannot achieve satisfactory performance when facing these lung nodules in various shapes or adhering to surrounding tissues. Enlightened by Dai et al. \cite{b22}, who present a network in which the convolutional layers can learn to add offsets on each sample location, we consider that it is necessary for the network to improve the block focus which can treat these hard lung nodules better. Compared to the spatial attention mechanism \cite{b27,b28}, deformable convolution does not depend on lots of training data limited by patient privacy and can accomplish feature extraction quickly. Hence, we intend to integrate deformable convolutional layers into our baseline model to enhance its focus. The convolution of each location $p_0$ in the CT image can be expressed as
\begin{equation}
y_1\left( p_0 \right) =\sum_{P_n\in R}{w\left( p_n \right) x\left( p_0+p_n \right)}\label{eq1}
\end{equation}
where $x$ represents the input feature map, $w$ represents the weight of sampled values, and $R$ defines the receptive field size and dilation. In the deformable convolution, $R$ is augmented with offsets, and $\bigtriangleup p_n$ represents the offsets. The deformable convolution of each location $p_0$ in the CT image can be expressed as
\begin{equation}
y_2\left( p_0 \right) =\sum_{P_n\in R}{w\left( p_n \right) x\left( p_0+p_n+\bigtriangleup p_n \right).}\label{eq2}
\end{equation}

\begin{figure}
\centerline{\includegraphics [width=0.35\textwidth] {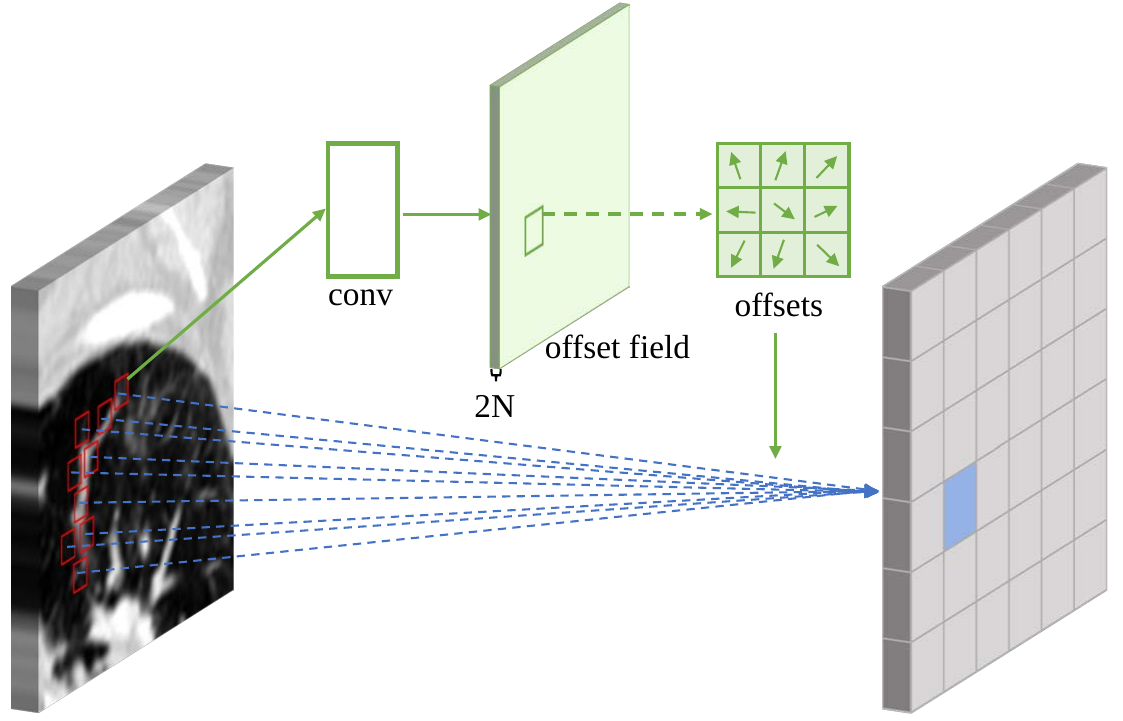}}
\caption{The deformable convolution operation.}
\label{fig2}
\end{figure}

The offsets on sampling location can be learned automatically during the training process. The operation of deformable convolution is shown in Fig. \ref{fig2}. Obviously, the shapes of receptive fields of the deformable convolution can better focus on the lung nodule area, which enables the ability to treat the hard lung nodules of our network. However, the replacement is not random. The deep layers usually extract some semantic information. Thus, we replace the front several convolutional layers with deformable ones, which will be introduced in detail in section \ref{ex}.

\subsection{Self-Paced Learning Strategy}
The multi-branch ensemble learning based on 3D CNN models greatly reduces false positive lung nodules \cite{b20}. However, the network could still produce inaccurate results when detecting hard lung nodules. The original concept of curriculum learning was first proposed by Bengio et al. \cite{b23} who trained the machine learning model with the easier subsets and then gradually increased the difficulty level of data until the whole training dataset. Similar to the learning process of human beings, the curriculum learning strategy could improve the model performance on detection tasks and accelerate the training process. Self-paced learning (SPL) \cite{b24,b25} is a primary branch of curriculum learning, and it can automate the difficulty measurer by taking the example training loss of the current model as criteria.

The loss of CNN can be expressed as
\begin{equation}
E\left( w \right) =\sum_{i=1}^n{L\left( y_i,g\left( x_i,w \right) \right)}\label{eq3}
\end{equation}
where $x_i$ denotes the $i$th observed sample, $y_i$ represents its label. $L\left( y_i,g\left( x_i,w \right) \right)$ denotes the loss function which calculates the cost between the ground truth label $y_i$ and the estimated label $g\left( x_i,w \right)$. $w$ represents the model parameter inside the decision function $g$. We introduce a variable $v$ that indicates whether the $i$th sample is easy or not. The SPL is able to learn the model parameter $w$ and the latent weight $v$ by minimizing 
\begin{equation}
\underset{w,v}{\min}\mathbb{E} \left( w,v \right) =\sum_{i=1}^n{v_iL\left( y_i,g\left( x_i,w \right) \right) +\lambda ^tf\left( v;t \right)}\label{eq4}
\end{equation}
where $v=\left[ 0,1 \right] ^n$, $f(v;t)$ is a proposed dynamic self-paced function with respect
to the weight $v$ and the time $t$, the parameter $\lambda$ is introduced to control the learning pace. In the process of the SPL calculation, we gradually increase $\lambda$ to learn new samples. When $\lambda$ is small, only easy samples with small losses will be considered into training. As $\lambda$ grows, hard samples with larger losses will be gradually added to train a more “mature” model. $f(v;t)$ can be described as follows
\begin{equation}
f\left( v;t \right) =\frac{1}{q\left( t \right)}\left\| v \right\| _{2}^{q\left( t \right)}-\sum_{i=1}^n{v_i}\label{eq5}
\end{equation}
where $q(t)>1$ and $q(t)$ is a monotonic decreasing function with respect to $t$.
\begin{figure}
\centering
\includegraphics [width=0.95\textwidth] {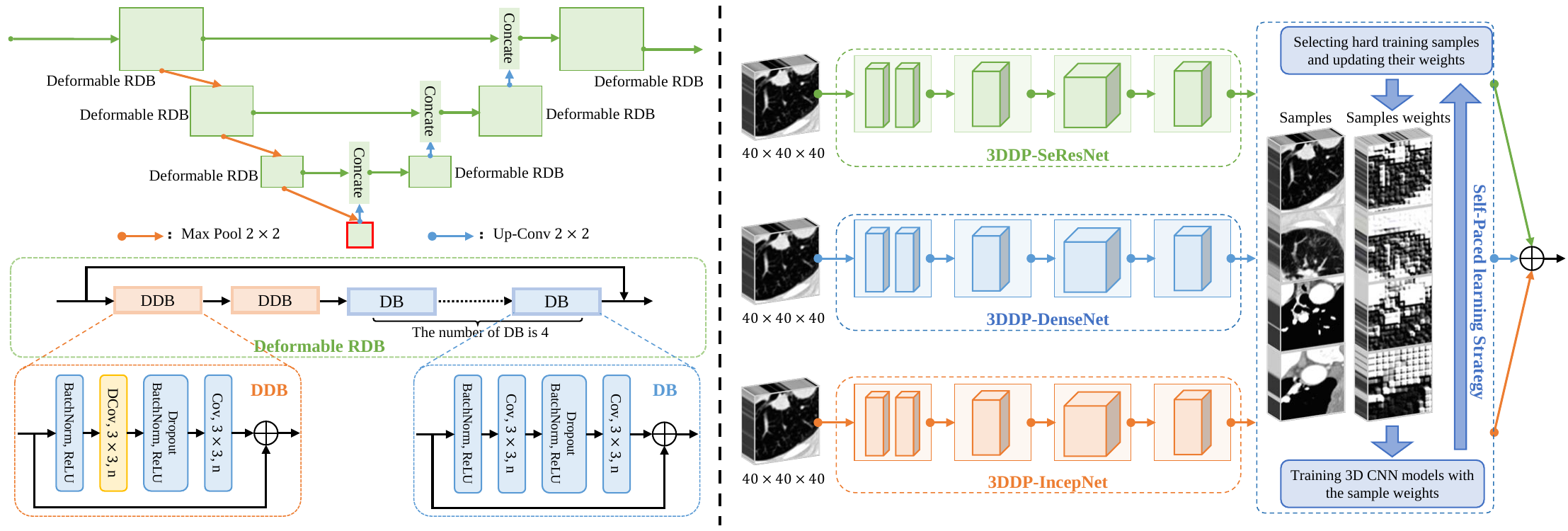}
\caption{The pipeline of our method. (a) shows the architecture of the improved U-Net, and (b) shows the multi-branch ensemble learning architecture based on three kinds of 3D CNN models that use the dual pooling approach. The CT images are processed by the improved U-Net that can detect all lung nodule candidates, and then the multi-branch ensemble learning 3D CNN utilizes the Self-Paced Learning strategy to reduce the false positive lung nodules. In the figure, "Max Pool" means an adaptive max pooling layer with the output size, "Up-Conv" means a standard convolutional layer with the kernel size, "DConv" means a deformable convolutional layer with the kernel size and output channel number, "BatchNorm" means batch normalization operation, "DB" defines a dense block, "DDB" defines a deformable dense block, "Deformable RDB" defines a deformable residual dense block, "3DDP-DenseNet" defines a 3D-DenseNet based on dual pooling (DP) and the other two are similarly defined.}
\Description{The pipeline of our method.}
\label{fig3}
\end{figure}

\subsection{Network Architecture}
The architecture of our modified network is depicted in Fig. \ref{fig3}. In order to enhance focus on the lung nodule area in CT images, deformable RDB consisting of 2 DDBs and 4 DBs is utilized in the improved U-Net architecture. Each DDB consists of a $3\times3$ deformable convolution and a $3\times3$ convolution with a skip connection. At the same time, the self-paced learning strategy is adopted in the multi-branch ensemble learning architecture to pay more attention to the hard lung nodules in all CT images. The multi-branch ensemble learning architecture automatically learns the sample weights according to the training loss. 

\section{Experiments}\label{ex}
Below, we will explain the implementation details of our method, including the data, evaluation criterion, and experimental results.

\subsection{Datasets}
We evaluate the proposed framework on the LUNA16 dataset \cite{b21}, which contains 888 CT images with the location centroids and diameters of the pulmonary nodules annotated. The dataset is a subset of the LIDC-IDRI dataset containing 2610 lung nodules. Each of them was labeled by up to four experienced radiologists. Moreover, each radiologist classifies the identified lesions into three categories: non-nodular (other tissues or background), nodules larger than 3 mm in diameter, and nodules less than 3 mm in diameter. Nodules that are larger than 3 mm in diameter, marked by three or four radiologists, are used as the gold standard, and nodules that are less than 3 mm in diameter and marked by only one or two radiologists will be ignored. We randomly divide the dataset into ten subsets for ten-fold cross validation.

\subsection{Evaluation Criterion}
The evaluation criterion is the same as those used in the LUNA16 competition. The CPM (competition performance metric) is defined as the average sensitivity at seven predefined false positive rates (these values are 0.125, 0.25, 0.5, 1, 2, 4, 8) to assess the performance of our method, and it can be calculated as
\begin{equation}
CPM=\frac{1}{N}\sum_{i\in C}{\mathrm{Recall}_{fpr=i}}\label{eq7}
\end{equation}
where the value of $N$ is 7, $C$ is $\left\{ 0.125,0.25,0.5,1,2,4,8 \right\}$, $fpr$ is the average number of false positives per scan, and ${Recall}_{fpr=i}$ represents the recall rate corresponding to $fpr = i$.

\subsection{Implementation Details}

In the first stage, we use Adam \cite{b29} optimizer to update the parameters. In order to prevent over-fitting, we adopt the early termination strategy. When the performance does not improve, the training will continue for 5 final epochs. In our experiment, the total number of training epochs is 15. The initial learning rate is 0.0001, the learning rate is 0.00001 for the fine-tuning, and the batch size is set to 64.

In the second stage, the multi-branch ensemble learning architecture consists of 3DDP-SeResNet,  3DDP-DenseNet, and 3DDP-IncepNet. The training process of 3D CNN networks for false positive reduction is identical. The SGD optimizer is used to update the model parameters, and the corresponding initial learning rate is 0.001. After that, each generation of learning rate decays to 0.9 of the previous generation. The batch size is set to 64. The three networks are trained for 20 epochs. Training the 3DDP-SeResNet,  3DDP-DenseNet, and 3DDP-IncepNet costs 28, 25, and 23 hours, respectively.

The proposed method is implemented by PyTorch. All expriments are carried out on 8 Nvidia 2080Ti GPUs.

\subsection{Ablation Study}
\subsubsection{The Impact of Proposed Deformable DB}

To validate the effectiveness of Deformable DB in our method, we design a TSCNN-based ablation experiment. The original RDB in TSCNN consists of 6 DBs. We replace the front several DBs with the designed DDB. From Table \ref{tab1}, the last three rows verify the validity of the components. We can observe that the CPM of our model is 0.919, 0.929, 0.932, and 0.932 when replacing zero, the front one, the front two, and the front three DBs, respectively. After introducing more than 2 Deformable DBs, the CPM will no longer grow obviously. Due to the increasing parameters, the training of our model will consume more time. Thus, we can obtain the best performance at "1,2-DDB" considering the balance between the gain and the time consumption. 

\begin{table}
\caption{Ablation Experiment of DDB Structure}
\label{tab1}
\begin{tabular}{ccc}
\toprule
\textbf{Method} & \textbf{CPM}& \textbf{Training Time}\\
\midrule
Non-DDB & 0.919 & 45h\\
1-DDB & 0.929 & 48h\\
\textbf{1,2-DDB} &\textbf{0.932} & \textbf{50h}\\
1,2,3-DDB & 0.932 & 54h\\
\bottomrule
\end{tabular}
\end{table}

\subsubsection{The Validity of Self-Paced Learning}

To demonstrate that it is better to utilize the self-paced learning strategy, we train the original TSCNN model with an equal training strategy and then train that with the new strategy. The results of ablation experiments are shown in Table \ref{tab2}. In this table, it can be seen that the TSCNN with SPL has a significant impact on CPM even on each scan, which directly verifies the effectiveness of the proposed improvements.

\begin{table}
\caption{Ablation Experiment of Self-Paced Learning Strategy on Different False Positive Rates}
\label{tab2}
\begin{tabular}{ccccccccc}
\toprule
\textbf{Method} & \textbf{0.125} & \textbf{0.25} & \textbf{0.5} & \textbf{1} & \textbf{2} & \textbf{4} & \textbf{8} & \textbf{CPM} \\
\midrule
TSCNN & 0.843 & 0.891 & 0.921 & 0.932 & 0.941 & 0.951 & 0.957 & 0.919\\
TSCNN + SPL & 0.858 & 0.905 & 0.931 & 0.941 & 0.953 & 0.959 & 0.963 & 0.93 \\
\bottomrule
\end{tabular}
\end{table}

\subsection{Comparative Evaluation}

\begin{figure}
\centering
\includegraphics [width=0.65\textwidth] {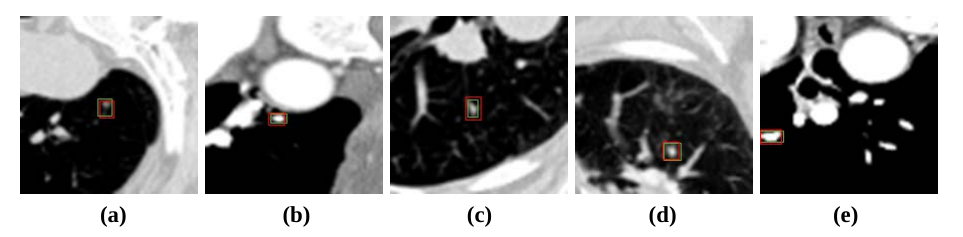}
\caption{The detection results of our method.}
\Description{The green box represents the label and the red box represents the prediction.}
\label{fig4}
\end{figure}

We compare our approach with other methods based on CNN. All the results are based on the LUNA16 dataset that is randomly divided into ten subsets for ten-fold cross validation. As shown in Table \ref{tab3}, it is obvious that our proposed method which improves the focus of the model is effective, and reaches satisfactory performance. Especially, the CPM increases by 0.9$\%$ compared with TSCNN. Our method has a great improvement at each scan which means that it could be more effective to detect those hard lung nodules. There are some detection results shown in Fig. \ref{fig4}.

\begin{table}
\caption{Comparison with Different Lung Nodule Detection Methods on LUNA16 Dataset}
\label{tab3}
\begin{tabular}{ccccccccc}
\toprule
\textbf{Method} & \textbf{0.125} & \textbf{0.25} & \textbf{0.5} & \textbf{1} & \textbf{2} & \textbf{4} & \textbf{8} & \textbf{CPM} \\
\midrule
Hamidian et al.\cite{b18} & 0.583 & 0.687 & 0.742 & 0.828 & 0.886 & 0.918 & 0.933 & 0.797 \\
Xie et al.\cite{b29} & 0.734 & 0.744 & 0.763 & 0.796 & 0.824 & 0.832 & 0.834 & 0.790\\
Dou et al.\cite{b30} &0.659 & 0.745 & 0.819 & 0.865 & 0.906 & 0.933 & 0.946 & 0.839\\
TSCNN\cite{b20} & 0.848 & 0.899 & 0.925 & 0.936 & 0.949 & 0.957 & 0.960 & 0.925\\
\textbf{Ours} & \textbf{0.862} & \textbf{0.911} & \textbf{0.934} & \textbf{0.944} & \textbf{0.958} & \textbf{0.963} & \textbf{0.967} & \textbf{0.934} \\
\bottomrule
\end{tabular}
\end{table}

\section{Conclusion}
 
For dealing with hard lung nodules in CT images, we consider improving feature extraction and introducing training strategies to enhance the model's focus on hard samples. We propose a method that introduces the deformable convolution to pay more attention to the areas containing lung nodules in each CT image. At the same time, the self-paced learning strategy was utilized to focus on those CT images that contain hard samples in the whole LUNA16 dataset. Our model based on TSCNN architecture has achieved a satisfactory performance. In the future, as more complicated detection networks are available and the attention mechanism can be involved, our goal is to design a CAD system that can deal with CT images in which the lung nodules are more difficult in small datasets.

\bibliographystyle{ACM-Reference-Format}
\bibliography{sample-base}

\end{document}